\documentclass[aps,pre,twocolumn,10pt,superscriptaddress,nofootinbib,balancelastpage]{revtex4-1} 
\usepackage[latin1]{inputenc}
\usepackage{amsmath,amssymb}
\usepackage{mathrsfs} 
\usepackage[capitalise]{cleveref}
\usepackage{siunitx}
\usepackage{braket}
\usepackage{calc}
\usepackage{tabularx}
\usepackage{dsfont}
\usepackage{color}
\usepackage{ifthen}

\allowdisplaybreaks

\newcommand{\ii}{\mathrm{i}}%
\newcommand{\dif}{\mathrm{d}}%
\newcommand{\norm}[1]{\lVert#1\rVert}%
\newcommand{\ZT}[1]{\textquotedblleft#1\textquotedblright}%

\newcolumntype{Y}{>{\centering\arraybackslash}X}%
\newcolumntype{Z}{>{\raggedright\arraybackslash}X}%

\newlength{\myl}%
\newcommand{\SUM}[2]{{\setlength{\myl}{\widthof{$\displaystyle\sum_{#1}^{#2}$}*\real{0.5}-\widthof{$\displaystyle\sum$}*\real{0.5}}\sum_{#1}^{#2}\;\hspace{-\the\myl}}}
\newcommand{\INT}[3]{\settowidth{\myl}{$\displaystyle\int_{#1}^{#2}$}{\int_{#1}^{#2}\;\;\;\hspace{-\the\myl}\dif #3}\,}
\newcommand{\TINT}[3]{\settowidth{\myl}{$\int_{#1}^{#2}$}{\int_{#1}^{#2}\!\ifthenelse{\equal{#1#2}{}}{}{\;\;\;\;\hspace{-\the\myl}}\dif #3}\,}%
\newcommand{\EINT}[3]{\settowidth{\myl}{$\int_{#1}^{#2}$}{\int_{#1}^{#2}\;\;\;\,\hspace{-\the\myl}\dif #3}\,}

\newcommand{\Clmn}[3]{c_{#1#2#3}}%
\newcommand{\CIIIDIJa}{c^{(3\mathrm{D})}}%
\newcommand{\CIIIDIJb}[2]{c^{(3\mathrm{D})}_{#1#2}}%
\newcommand{\CIIIDIJc}[4]{c^{(3\mathrm{D})}_{#1#2#3#4}}%
\newcommand{\CIIIDIJd}[6]{c^{(3\mathrm{D})}_{#1#2#3#4#5#6}}%
\newcommand{\WDvarpi}[3]{D^{#1}_{#2#3}}%
\newcommand{\EinrueckabstandI}{\quad\,}%
\newcommand{\EinrueckabstandII}{\quad\;}%

\newcommand{\uu}{\hat{u}}%
\newcommand{\INTOI}{\INT{S_{1}}{}{\Omega}}%
\newcommand{\TINTOI}{\TINT{S_{1}}{}{\Omega}}%
\newcommand{\INTOII}{\INT{S_{2}}{}{\Omega}}%
\newcommand{\TINTOII}{\TINT{S_{2}}{}{\Omega}}%
\newcommand{\ff}[3]{f_{#3}^{(#1\mathrm{D})}}%
\newcommand{\ffp}[3]{f_{#3}^{(#1\mathrm{D})'}}%
\newcommand{\TT}[3]{\mathrm{T}_{#3}^{(#1\mathrm{D})}}%

\begin{document}
	
\title{Relations between angular and Cartesian orientational expansions}
\author{Michael te Vrugt}
\affiliation{Institut f\"ur Theoretische Physik, Westf\"alische Wilhelms-Universit\"at M\"unster, D-48149 M\"unster, Germany}
\affiliation{Center for Soft Nanoscience, Westf\"alische Wilhelms-Universit\"at M\"unster, D-48149 M\"unster, Germany}
	
\author{Raphael Wittkowski}
\email[Corresponding author: ]{raphael.wittkowski@uni-muenster.de}
\affiliation{Institut f\"ur Theoretische Physik, Westf\"alische Wilhelms-Universit\"at M\"unster, D-48149 M\"unster, Germany}
\affiliation{Center for Soft Nanoscience, Westf\"alische Wilhelms-Universit\"at M\"unster, D-48149 M\"unster, Germany}	
\affiliation{Center for Nonlinear Science, Westf\"alische Wilhelms-Universit\"at M\"unster, D-48149 M\"unster, Germany}

\begin{abstract}		
Orientational expansions, which are widely used in the natural sciences, exist in angular and Cartesian form. Although these expansions are orderwise equivalent, it is difficult to relate them in practice. In this article, both types of expansions and their relations are explained in detail. We give explicit formulas for the conversion between angular and Cartesian expansion coefficients for functions depending on one, two, and three angles in two and three spatial dimensions. These formulas are useful, e.g., for comparing theoretical and experimental results in liquid crystal physics. The application of the expansions in the definition of orientational order parameters is also discussed.
\end{abstract}
\maketitle

\section{\label{introduction}Introduction}
Orientational expansions, i.e., expansions of the angular dependence of a function $f(\Omega)$, with $\Omega$ denoting an orientational variable, are used in many fields of the natural sciences, such as liquid crystal physics \cite{YuanMSTS2018,StenhammarWMC2016,WittmannMM2016,WittkowskiLB2010,WittkowskiLB2011,WittkowskiLB2011b}, active matter physics \cite{Pearce2019,MuellerYD2019,HartmannSPMSDDD2019,DoostmohammadiIYS2018,EllisPCGGF2018,PraetoriusVWL2018,OphausGT2018}, polymer physics \cite{WangLCWLLPSZH2017,WangLCWLLPSZH2017}, electrostatics \cite{SwartvDS2001,Applequist1989}, optics \cite{GrinterJ2018,NieminenRH2003,SheppardT1997}, geophysics \cite{WieczorekM2018}, astrophysics and cosmology \cite{BartoloKLRST2018,Kamionkowski2018}, general relativity \cite{Kundu1981,Zhang1986}, engineering \cite{ZuoSAD2019,BrinkmannW2018}, chemistry \cite{BannwarthEG2019,SinghAC2019}, and medicine \cite{LuYZT2019}. Important examples for orientational expansions are the Fourier expansion for $\Omega = \phi \in [0,2\pi)$, the expansion in spherical harmonics for $\Omega = (\theta,\phi) \in [0,\pi] \times [0,2\pi)$, and the expansion in outer products of a normalized orientation vector $\uu$ for $\Omega = \uu$.
The expansions can be classified in two main categories, which differ in the way the expansion coefficients transform under rotations: angular expansions (including the Fourier series and the spherical harmonics expansion) and Cartesian expansions (including expansions in symmetric traceless tensors and outer products of orientation vectors or rotation matrices\footnote{By \ZT{rotation matrix}, we always denote a matrix of the form \eqref{rotation2d} or \eqref{rotationmatrix} that rotates a Cartesian vector. Wigner rotation matrices, which correspond to the angular case, are referred to as \ZT{Wigner D-matrices}.}).

One of the main applications of such expansions is the description of the orientational order of liquid crystals. Here, we can use an orientation vector $\uu$ depending on the angle $\phi$ in two spatial dimensions (2D) or on two angles $\theta$ and $\phi$ in three spatial dimensions (3D) to specify the orientation of a particle. A system of particles is then described using a distribution function $f(\uu)$. The coefficients of the expansion of the function $f(\uu)$ provide orientational order parameters. Of particular importance are the Cartesian order parameters of first and second order, given by the polarization $\vec{P}$ and the nematic tensor $Q$, respectively \cite{WittkowskiSC2017}. They are used, e.g., to formulate field theories for liquid crystals \cite{deGennesP1995,EmmerichEtAl2012}. Also order parameters of third order are useful for the description of certain phase transitions \cite{LubenskyR2002}.

In three spatial dimensions, the description of the orientational order using one orientation vector is only sufficient if the particles have an axis of continuous rotational symmetry (\ZT{uniaxial particles}\footnote{In this work, we use the terminology \ZT{uniaxial} and \ZT{biaxial} to distinguish particles with and without an axis of continuous rotational symmetry, respectively. Note that these adjectives can also be used in a different way, referring to optical properties of a system \cite{Rosso2007}. Moreover, we make the common approximation of assuming all particles to be rigid bodies.}). 
The situation is more complex if one considers particles without such a symmetry (\ZT{biaxial particles}). Here, a description of their orientation requires two orientation vectors or, alternatively, three angles such as the Euler angles $(\theta,\phi,\chi) \in [0,\pi] \times [0,2\pi) \times [0,2\pi)$, which correspond to a rotation $R$ that maps the laboratory-fixed Cartesian coordinate system onto a body-fixed one \cite{Rosso2007}. Therefore, the definition of order parameters for biaxial particles requires more general orientational expansions \cite{Rosso2007,MatsuyamaAWF2019,LuckhurstS2015,EhrentrautM1998,Turzi2011}. Based on liquid crystal terminology, we will refer to expansions for functions $f(\uu)$ as \ZT{uniaxial} and to expansions for functions $f(R)$ as \ZT{biaxial}. Note, however, that the applicability of our results is not restricted to liquid crystal physics, but extends to all fields where such expansions are used.

Both angular and Cartesian expansions have their own advantages and disadvantages. Angular expansions allow to make use of the mathematical properties of circular or spherical harmonics and have a smaller number of expansion coefficients, since they are all independent. Cartesian expansions, on the other hand, have a clearer geometrical interpretation and are more directly connected to computer simulations and experiments \cite{Turzi2011}. A good example is the study of the orientational order of liquid crystals consisting of low-symmetry molecules. For theoretical studies, order parameters based on an expansion in Wigner D-matrices, which have a useful and well-known mathematical structure, are widely used \cite{Blum2012,GrayG1984}. Experimentalists, on the other hand, frequently use the Saup\'{e} ordering matrix -- a Cartesian order parameter -- which is more practical in describing outcomes of, e.g., nuclear magnetic resonance (NMR) experiments \cite{Teng2007}. This can constitute a difficulty in comparing the results of theoretical calculations with computer simulations and experiments.

The problem can be solved if the relations between the different types of expansions are known. It is very difficult to give such an expression in a general form \cite{JoslinG1984}. One can, however, explicitly calculate these relations for the lowest orders, which is sufficient for almost all practical applications. In this work, we provide tables containing all relevant relations. This includes coefficients of zeroth to third order for functions depending on one angle (2D, uniaxial), two angles (3D, uniaxial), and three angles (3D, biaxial). These relations, which can be found in an Appendix, constitute the main result of this article. 
The conversion rules can be used in any field of the natural sciences where orientational distributions are relevant. For example, they allow to easily convert Cartesian data from a computer simulation or an NMR experiment on a liquid crystal into a form that can be compared with theoretical calculations that use angular functions. Moreover, these equations allow, e.g., to easily calculate the dipole vector corresponding to data that are given in the form of an expansion in spherical harmonics.

In addition, we give an overview over the mathematical structure of uniaxial and biaxial expansions, including definitions of all special functions that are involved. We also provide formulas for the expansion coefficients. Such an overview is very useful and difficult to find in the literature. It also clarifies the conventions used for obtaining the conversion formulas. Furthermore, we explain how these expansions allow to define order parameters for liquid crystals. 

This article is structured as follows: In \cref{uniaxial}, we describe angular and Cartesian expansions with the relevant functions, coefficients, and order parameters for the uniaxial case. The biaxial expansions are described in \cref{biaxial}. We summarize the work and give an outlook in \cref{conclusion}. The relations between the uniaxial expansion coefficients are listed in \cref{relation} and those for the biaxial ones in \cref{relations}. A list of elements of the Wigner D-matrices can be found in \cref{list}.

\section{\label{uniaxial}Uniaxial expansions}
We consider uniaxial particles in two and three spatial dimensions. 
The orientational distribution of the particles is described by a scalar orientation-dependent function $f$.  
In a 2D system, this function can be parameterized as $f(\phi)\equiv f(\uu)$ with the polar angle $\phi\in[0,2\pi)$ and the orientational unit vector $\uu(\phi)=(\cos(\phi),\sin(\phi))^{\mathrm{T}}$, where the superscript $^{\mathrm{T}}$ denotes a transposition. When considering a 3D system, the function can be parameterized as $f(\theta,\phi)\equiv f(\uu)$ with the spherical coordinates $\theta\in[0,\pi]$ and $\phi\in[0,2\pi)$ as well as the orientational unit vector $\uu(\theta,\phi)=(\sin(\theta)\cos(\phi),\sin(\theta)\sin(\phi),\cos(\theta))^{\mathrm{T}}$.

\subsection{Angular multipole expansion}
The scalar orientation-dependent function $f$ can orthogonally be expanded in terms of angular coordinates $\phi$ (in 2D) and $\theta$ and $\phi$ (in 3D). We follow Refs.\ \cite{JoslinG1983,GrayG1984}.
	
\subsubsection{Circular multipole expansion (2D)}
In the case of two spatial dimensions, the angular multipole expansion is also called \ZT{circular multipole expansion} and identical to the Fourier series expansion
\begin{equation}
f(\phi)=\SUM{k=-\infty}{\infty} f_{k} e^{\ii k \phi} 
\label{fphi}%
\end{equation}
with the imaginary unit $\ii$ and the circular harmonics $e^{\ii k \phi}$. The corresponding (Fourier) expansion coefficients are given by 
\begin{equation}
f_{k}=\frac{1}{2\pi} \INTOI f(\phi) e^{-\ii k \phi}, 
\label{fk}%
\end{equation}
where $\TINTOI=\TINT{0}{2\pi}{\phi}$ denotes an angular integration over the unit circle $S_{1}$. 
In general, these expansion coefficients are independent.
If the coordinate system is rotated by an angle $\varphi$, the expansion coefficients $f_k$ change to
\begin{equation}
f'_{k}=f_k e^{\ii k \varphi}. 
\label{rotationfk}%
\end{equation}

\subsubsection{Spherical multipole expansion (3D)}
When there are three spatial dimensions, the angular multipole expansion is identical to the \ZT{spherical multipole expansion} 
\begin{equation}
f(\theta,\phi)=\SUM{l=0}{\infty}\SUM{m=-l}{l} f_{lm} Y_{lm}(\theta,\phi)
\label{fthetaphi}%
\end{equation}
with the spherical harmonics
\begin{equation}
Y_{lm}(\theta,\phi)=\sqrt{\frac{2l+1}{4\pi} \frac{(l-m)!}{(l+m)!}} P_{lm}(\cos(\theta))e^{\ii m \phi}
\label{sh}%
\end{equation}
and the associated Legendre polynomials
\begin{equation}
P_{lm}(x)=\frac{(-1)^{m}}{2^{l}l!} (1-x^{2})^{m/2} \partial_{x}^{l+m} (x^{2}-1)^{l}.
\end{equation}
The latter two functions are stated here explicitly to avoid confusion with other conventions. 
Now, the expansion coefficients are given by 
\begin{equation}
f_{lm}=\INTOII f(\theta,\phi) Y^\star_{lm}(\theta,\phi),
\label{flm}%
\end{equation}
where $\TINTOII=\TINT{0}{\pi}{\theta}\sin(\theta)\TINT{0}{2\pi}{\phi}$ is an angular integration over the unit sphere $S_{2}$
and the star $\star$ denotes complex conjugation. 
As is the 2D case, the expansion coefficients are in general independent. 
Under passive\footnote{\ZT{Active} and \ZT{passive} here refer to two different conventions used in the description of rotations, whereas elsewhere in the article we use this terminology to distinguish particles with and without self-propulsion, respectively. An active rotation corresponds to a rotation of a body in a fixed coordinate system, whereas in a passive rotation the coordinate axes are rotated. Switching between these conventions corresponds to turning a clockwise into a counterclockwise rotation \cite{GrayG1984}, which is why it is important to clarify the convention. Throughout this article, we follow the conventions used by Gray and Gubbins \cite{GrayG1984} for the description of rotations.} rotations, the expansion coefficients transform according to 
\begin{equation}
f_{lm} = \SUM{n=-l}{l}D^l_{mn}(\vec{\varpi})f_{ln}',
\label{flns}%
\end{equation}
where the $D^l_{mn}(\vec{\varpi})$ are the Wigner D-matrices depending on the Euler angles $\vec{\varpi}$ (see \cref{dlm}).

\subsection{\label{cartesianuniaxial}Cartesian multipole expansion}
The scalar orientation-dependent function $f$ can also orthogonally be expanded in terms of the orientation vector $\uu$. 
This applies to both 2D and 3D. Our presentation of this expansion follows Refs.\ \cite{JoslinG1983} (for 2D) and \cite{GrayG1984} (for 3D). 
	
The so-called \ZT{Cartesian multipole expansion} is given by 
\begin{equation}
f(\uu)=\SUM{l=0}{\infty} \SUM{i_{1},\dotsc,i_{l}=1}{d} \ff{d}{l}{i_{1}\dotsb i_{l}} u_{i_{1}}\!\dotsb u_{i_{l}},
\label{fu}%
\end{equation}
where $d\in\{2,3\}$ is the number of spatial dimensions and $u_{i}$ is the $i$-th element of the orientation vector $\uu=(u_{1},\dotsc,u_{d})^{\mathrm{T}}$.
For this expansion, the corresponding expansion coefficients are obtained as 
\begin{equation}
\ff{d}{l}{i_{1}\dotsb i_{l}} = A^{(d\mathrm{D})}_{l} \INT{S_{d-1}}{}{\Omega} f(\uu) \TT{d}{l}{i_{1}\dotsb i_{l}}
\label{eq:fdD}%
\end{equation}
with the prefactors 
\begin{equation}
A^{(2\mathrm{D})}_{l} = \frac{2-\delta_{0l}}{\Omega_{2}}, \qquad
A^{(3\mathrm{D})}_{l} = \frac{2l+1}{\Omega_{3}}
\label{prefactors}%
\end{equation}
and the angular normalization factors 
\begin{equation}
\Omega_{d} = \INT{S_{d-1}}{}{\Omega}\! 
= \begin{cases}
2\pi, \quad \text{ for }d=2,\\
4\pi, \quad \text{ for }d=3.
\end{cases}
\label{normalizationfactors}%
\end{equation}
The tensors $\TT{d}{l}{i_{1}\dotsb i_{l}}$ on the right-hand side of Eq.\ \eqref{eq:fdD} equal the tensor Chebyshev polynomials of the first kind $\TT{2}{l}{i_{1}\dotsb i_{l}}$ for $d=2$ and the tensor Legendre polynomials $\TT{3}{l}{i_{1}\dotsb i_{l}}$ for $d=3$. They are given by 
\begin{align}
\TT{2}{l}{i_{1}\dotsb i_{l}} &= \frac{(-1)^{l}}{l!} (l+\delta_{0l}) \partial_{i_{1}}\!\dotsb\partial_{i_{l}}(1-\ln(r))\bigg\rvert_{\vec{r}=\uu}, \\
\TT{3}{l}{i_{1}\dotsb i_{l}} &= \frac{(-1)^{l}}{l!} \partial_{i_{1}}\!\dotsb\partial_{i_{l}}\frac{1}{r}\bigg\rvert_{\vec{r}=\uu}
\label{tensorlegendre}%
\end{align}
with $r=\norm{\vec{r}}$ and the Euclidean norm $\norm{\cdot}$. In Table \ref{tab:T}, the first four of these tensors for $d=2$ and $d=3$ are listed explicitly.
\begin{table}[htbp]
\centering\renewcommand*{\arraystretch}{1.2}
\begin{tabular}{ccc}
\hline
$\boldsymbol{l}$ & $\boldsymbol{\TT{2}{l}{i_{1}\dotsb i_{l}}}$ & $\boldsymbol{\TT{3}{l}{i_{1}\dotsb i_{l}}}$ \\\hline 
$0$ & $1$ & $1$ \\\hline 
$1$ & $u_{i_{1}}$ & $u_{i_{1}}$ \\\hline 
$2$ & $2u_{i_{1}}u_{i_{2}}-\delta_{i_{1}i_{2}}$ & $\frac{1}{2} (3u_{i_{1}}u_{i_{2}}-\delta_{i_{1}i_{2}})$ \\\hline 
$3$ & $4u_{i_{1}}u_{i_{2}}u_{i_{3}}-3(u_{i_{1}}\delta_{i_{2}i_{3}})^{\mathrm{sym}}$ & $\frac{1}{2}(5u_{i_{1}}u_{i_{2}}u_{i_{3}}-3(u_{i_{1}}\delta_{i_{2}i_{3}})^{\mathrm{sym}})$ \\\hline 
\raisebox{-0.6mm}[\ht\strutbox]{$\vdots$} & \raisebox{-0.6mm}[\ht\strutbox]{$\vdots$} & \raisebox{-0.6mm}[\ht\strutbox]{$\vdots$} \\\hline 
\end{tabular} 
\caption{\label{tab:T}Tensor Chebyshev polynomials of the first kind $\TT{2}{l}{i_{1}\dotsb i_{l}}$ and tensor Legendre polynomials $\TT{3}{l}{i_{1}\dotsb i_{l}}$ for different orders $l$, where $(\cdot)^{\mathrm{sym}}$ denotes the symmetrization of a tensor.}
\end{table}
The tensors $\TT{d}{l}{i_{1}\dotsb i_{l}}$ and therefore also the Cartesian coefficient tensors \eqref{eq:fdD} are symmetric and traceless for $l>1$. 
When $f(\uu)$ is real, the same applies to the Cartesian coefficient tensors $\ff{d}{l}{i_{1}\dotsb i_{l}}$.
In general, not more than $2-\delta_{0l}$ (in 2D) and $2l+1$ (in 3D) elements of a Cartesian coefficient tensor of order $l$ can be independent. 
The first four Cartesian coefficient tensors for $d=2$ and $d=3$ are listed explicitly in Table \ref{tab:f}.
\begin{table*}[htbp]
\centering\renewcommand*{\arraystretch}{1.2}
\begin{tabularx}{\linewidth}{cYY}
\hline
$\boldsymbol{l}$ & $\boldsymbol{\ff{2}{l}{i_{1}\dotsb i_{l}}}$ & $\boldsymbol{\ff{3}{l}{i_{1}\dotsb i_{l}}}$ \\\hline 
$0$ & $\frac{1}{2\pi}\TINTOI f(\uu)$ & $\frac{1}{4\pi}\TINTOII f(\uu)$ \\\hline 
$1$ & $\frac{1}{\pi}\TINTOI f(\uu) u_{i_{1}}$ & $\frac{3}{4\pi}\TINTOII f(\uu) u_{i_{1}}$ \\\hline 
$2$ & $\frac{1}{\pi}\TINTOI f(\uu) (2u_{i_{1}}u_{i_{2}}-\delta_{i_{1}i_{2}})$ & $\frac{5}{8\pi}\TINTOII f(\uu) (3u_{i_{1}}u_{i_{2}}-\delta_{i_{1}i_{2}})$ \\\hline 
$3$ & $\frac{1}{\pi}\TINTOI f(\uu)(4u_{i_{1}}u_{i_{2}}u_{i_{3}}-u_{i_{1}}\delta_{i_{2}i_{3}}-u_{i_{2}}\delta_{i_{3}i_{1}}-u_{i_{3}}\delta_{i_{1}i_{2}})$ & $\frac{7}{8\pi}\TINTOII f(\uu) (5u_{i_{1}}u_{i_{2}}u_{i_{3}}-u_{i_{1}}\delta_{i_{2}i_{3}}-u_{i_{2}}\delta_{i_{3}i_{1}}-u_{i_{3}}\delta_{i_{1}i_{2}})$ \\\hline 
\raisebox{-0.6mm}[\ht\strutbox]{$\vdots$} & \raisebox{-0.6mm}[\ht\strutbox]{$\vdots$} & \raisebox{-0.6mm}[\ht\strutbox]{$\vdots$} \\\hline 
\end{tabularx} 
\caption{\label{tab:f}Cartesian expansion coefficients $\ff{2}{l}{i_{1}\dotsb i_{l}}$ and $\ff{3}{l}{i_{1}\dotsb i_{l}}$ for different orders $l$.}
\end{table*}
Under rotations, the expansion coefficients transform as Cartesian tensors, i.e., 
\begin{equation}
\ff{d}{l}{i_{1}\dotsb i_{l}} = \SUM{j_{1},\dotsc,j_{l}=1}{d}R_{i_1j_1} \!\dotsb R_{i_lj_l}\ffp{d}{l}{j_{1}\dotsb j_{l}}
\label{fj1jl}%
\end{equation}
with the rotation matrix $R_{ij}$ (see \cref{cbpcme}). The rotation matrix for $d=2$ is defined as
\begin{equation}
R(\phi) =
\begin{pmatrix}
\cos(\phi) & -\sin(\phi)\\
\sin(\phi) & \cos(\phi)
\end{pmatrix}
\label{rotation2d}
\end{equation}
and the rotation matrix for $d=3$ is given by \cref{rotationmatrix}.
	
An advantage of the Cartesian multipole expansion is that it is a direct expansion in the variable $\uu$, whereas the angular variables $\phi$ (in 2D) or $\theta$ and $\phi$ (in 3D) appear not directly but via exponential and trigonometric functions in the angular multipole expansion. On the other hand, the number of expansion coefficients is higher for the Cartesian multipole expansion although not more of the expansion coefficients can be independent. 
Despite of the differences of both types of expansions, they are equivalent. Moreover, each order of one expansion is equivalent to the same order of the other expansion.
This allows an orderwise mapping between both types of expansions and explains the maximal number of independent expansion coefficients for the Cartesian multipole expansion.  
In \cref{relation}, explicit equations expressing the expansion coefficients of an angular multipole expansion in terms of the expansion coefficients of a Cartesian multipole expansion and vice versa are given for dimensionalities $d=2$ and $d=3$ and up to third order. 

The equations for the expansion coefficients $f_{k}$ in terms of the expansion coefficients $\ff{2}{l}{i_{1}\dotsb i_{l}}$ have been obtained by inserting \cref{fu} into \cref{fk} and performing the angular integration. Similarly, we have expressed the $f_{lm}$ in terms of the $\ff{3}{l}{i_{1}\dotsb i_{l}}$ by inserting \cref{fu} into \cref{flm} and integrating. Finally, the $\ff{d}{l}{i_{1}\dotsb i_{l}}$ have been expressed in terms of the $f_{k}$ (in 2D) and $f_{lm}$ (in 3D) by inserting \cref{fphi} (in 2D) and \cref{fthetaphi} (in 3D) into \cref{eq:fdD} and evaluating the integrals.

\subsection{Order parameters}
The Cartesian coefficient tensors \eqref{eq:fdD} can be used as order parameters in liquid-crystal theory.
They correspond to the mean particle density ($l=0$, monopole moment), polarization vector ($l=1$, dipole moment), nematic tensor ($l=2$, quadrupole moment), tetratic tensor ($l=3$, octupole moment), and so on \cite{deGennesP1995,LubenskyR2002}. Usually, only order parameters up to second order are considered. 
	
In general, a liquid crystal consisting of particles that are (considered as) uniaxial is described microscopically using a probability distribution $f(\vec{r},\uu)$ that depends on position $\vec{r}$ and orientation $\uu$. The vector $\uu$ specifies the geometrical orientation of a particle or, for an active particle, the direction of self-propulsion \cite{BickmannW2019,BickmannW2019b}. A Cartesian expansion in the form \eqref{fu} up to second order gives
\begin{equation}
f(\vec{r},\uu) = \rho(\vec{r}) + \sum_{i=1}^{d}P_i(\vec{r})u_i + \sum_{i,j=1}^{d}Q_{ij}(\vec{r})u_i u_j 
\end{equation}
with for $d=2$
\begin{align}
\rho(\vec{r}) &= \frac{1}{2\pi}\INT{S_1}{}{\Omega}f(\vec{r},\uu),\\
P_i(\vec{r}) &= \frac{1}{\pi}\INT{S_1}{}{\Omega}f(\vec{r},\uu) u_i,\\
Q_{ij}(\vec{r}) &= \frac{2}{\pi}\INT{S_1}{}{\Omega}f(\vec{r},\uu) \Big(u_i u_j - \frac{1}{2}\delta_{ij}\Big)
\end{align} 
and for $d=3$
\begin{align}
\rho(\vec{r}) &= \frac{1}{4\pi}\INT{S_2}{}{\Omega}f(\vec{r},\uu),\\
P_i(\vec{r}) &= \frac{3}{4\pi}\INT{S_2}{}{\Omega}f(\vec{r},\uu) u_i,\\
Q_{ij}(\vec{r}) &= \frac{15}{8\pi}\INT{S_2}{}{\Omega}f(\vec{r},\uu) \Big(u_i u_j - \frac{1}{3}\delta_{ij}\Big).
\end{align}
Now suppose that we have a system of $N$ particles and that we can assign every particle a position $\vec{r}_n$ and an orientation vector $\uu_{n}=\uu(\phi_n)$ for $d=2$ and $\uu_{n}=\uu(\theta_n,\phi_n)$ for $d=3$ with $n\in\{1,\dotsc,N\}$.
In this case, we can write the microscopic one-particle distribution function as
\begin{equation}
f(\vec{r},\uu) = \sum_{i=1}^{N}\delta(\vec{r}-\vec{r}_n)\delta(\uu - \uu_n),
\label{micro}
\end{equation}
which gives the order parameters for $d=2$
\begin{align}
\rho(\vec{r}) &=\frac{1}{2\pi}\SUM{n=1}{N}\delta(\vec{r}-\vec{r}_n),\label{rho2D}\\
P_{i}(\vec{r}) &= \frac{1}{\pi}\SUM{n=1}{N}u_{n,i}\delta(\vec{r}-\vec{r}_n),\\
Q_{ij}(\vec{r}) &= \frac{2}{\pi}\SUM{n=1}{N}\Big(u_{n,i}u_{n,j} - \frac{1}{2}\delta_{ij}\Big)\delta(\vec{r}-\vec{r}_n)
\label{qij2D}%
\end{align}
and for $d=3$
\begin{align}
\rho(\vec{r}) &=\frac{1}{4\pi}\SUM{n=1}{N}\delta(\vec{r}-\vec{r}_n),\label{rho3D}\\
P_{i}(\vec{r}) &= \frac{3}{4\pi}\SUM{n=1}{N}u_{n,i}\delta(\vec{r}-\vec{r}_n),\\
Q_{ij}(\vec{r}) &= \frac{15}{8\pi}\SUM{n=1}{N}\Big(u_{n,i}u_{n,j} - \frac{1}{3}\delta_{ij}\Big)\delta(\vec{r}-\vec{r}_n)
\label{qij3D}%
\end{align}
with $u_{n,i}=(\uu_{n})_{i}$ denoting the $i$-th element of the vector $\uu_n$. 
These order parameters are, in the case of Eqs.\ \eqref{rho2D} and \eqref{rho3D} up to a prefactor, the usual microscopic definitions of the one-particle density $2(d-1)\pi\rho(\vec{r})$, polarization vector $\vec{P}(\vec{r})$, and nematic tensor $Q(\vec{r})$. They can be used, e.g., to describe a system of active particles where $\uu_n$ denotes the direction of self-propulsion of the $n$-th particle. In the case of passive apolar particles such as rods, where there is no physical difference between $\uu_n$ and $-\uu_n$, one can incorporate the head-tail symmetry by replacing $\delta(\uu - \uu_n)$ by $(\delta(\uu - \uu_n) + \delta(\uu + \uu_n))/2$ in \cref{micro}.

\section{\label{biaxial}Biaxial expansions}
We consider biaxial particles in three spatial dimensions.\footnote{In two spatial dimensions, one orientation vector is always sufficient, hence we do not need to consider the 2D case separately in this section.}  
The orientational distribution function of the particles can now be parameterized as $f(\theta,\phi,\chi)\equiv f(R)$ with the third Euler angle $\chi\in[0,2\pi)$ and the rotation $R \in$ SO(3) that can be represented by a rotation matrix $R_{ij}(\theta,\phi,\chi)$.

\subsection{\label{dlm}Angular multipole expansion}
For describing the orientation of a biaxial particle, we use the Euler angles $(\theta,\phi,\chi) \in [0,\pi] \times [0,2\pi) \times [0,2\pi)$ and introduce the vector $\vec{\varpi}=(\theta,\phi,\chi)^{\mathrm{T}}$ as a shorthand notation. The Euler angles can be defined in various ways. For convenience, we use the popular definition of Gray and Gubbins \cite{GrayG1984}. There, the angles $\theta$ and $\phi$ correspond to the usual angles of spherical coordinates, while the third angle $\chi \in [0,2\pi)$ describes a rotation about the axis defined by $\theta$ and $\phi$. The advantage of this convention is that, since the first two angles have the same definition as in the usual case of spherical coordinates, the order parameters for biaxial particles will contain the common definitions for uniaxial particles as a limiting case. 

The Euler angles are also a way to specify a rotation in three spatial dimensions, so that a function $f(\vec{\varpi})\equiv f(\theta,\phi,\chi)$ can be thought of as a function $f(R)$ with $R$ being a rotation represented by a $3 \times 3$ matrix. Thus, we essentially need to find a way for expanding a function that is defined on the rotation group SO(3). The Wigner D-matrices \cite{GrayG1984}
\begin{equation}
D^l_{mn}(\vec{\varpi}) = e^{-\ii m \phi}d^l_{mn}(\theta)e^{-\ii n \chi}
\label{eq:WignerDmatrices}%
\end{equation}
with the Wigner d-matrices
\begin{equation}
\begin{split}
d^l_{mn}(\theta)&=\sqrt{(l+m)!(l-m)!(l+n)!(l-n)!}\\
&\quad\; \sum_{k\in\mathcal{I}^{l}_{mn}}\frac{(-1)^k \cos(\frac{\theta}{2})^{2l+m-n-2k} \sin(\frac{\theta}{2})^{2k-m+n}}{(l+m-k)!(l-n-k)!k!(k-m+n)!}
\end{split}\label{dlmn}\raisetag{4.5em}%
\end{equation}
and $-l \leq m,n \leq l$ are irreducible representations of the group SO(3). The set $\mathcal{I}^{l}_{mn}$ contains all integers that $k$ can attain such that all factorial arguments in Eq.\ \eqref{dlmn} are nonnegative \cite{VarshalovichMK1988}. See Ref.\ \cite{VarshalovichMK1988} for differential and integral representations of the functions $d^l_{mn}(\theta)$. The functions $D^l_{mn}(\vec{\varpi})$ are also referred to as Wigner D-functions \cite{VarshalovichMK1988}, Wigner rotation matrices \cite{Tajima2015}, and four-dimensional spherical harmonics \cite{GrayG1984}. Note that the definition of the Wigner matrices used here assumes passive rotations \cite{GrayG1984,Man2016}.
	
An expansion for functions defined on SO(3) can be performed using the Peter-Weyl theorem, which states (roughly) that for a compact topological group $G$, a square-integrable function $f\in L^2(G)$ can be expanded in terms of matrix elements of the irreducible representations of $G$ \cite{Turzi2011}. Since these representations are given by the Wigner D-matrices for the group SO(3), we can expand a function $f(\vec{\varpi}) \in L^2(\mathrm{SO(3)})$ in the form \cite{GrayG1984,Turzi2011} 
\begin{equation}
f(\vec{\varpi})=\sum_{l=0}^{\infty}\sum_{m=-l}^{l}\sum_{n=-l}^{l}c_{lmn}D^l_{mn}(\vec{\varpi})
\label{dexp}%
\end{equation}
with the expansion coefficients
\begin{equation}
c_{lmn} = \frac{2l+1}{8\pi^2}\INT{\mathrm{SO(3)}}{}{\Omega} D^l_{mn}(\vec{\varpi})^\star f(\vec{\varpi}),
\label{clmn}%
\end{equation}
where the integral is defined as \cite{GrayG1984}
\begin{equation}
\INT{\mathrm{SO(3)}}{}{\Omega} = \INT{0}{2\pi}{\chi}\INT{0}{2\pi}{\phi}\INT{0}{\pi}{\theta}\sin(\theta).
\end{equation}
The expansion coefficients \eqref{clmn} are in general independent, such that there are $(2l+1)^2$ independent coefficients at order $l$.
	
Explicit expressions for the elements of the Wigner D-matrices up to order $l=3$ are given in \cref{list}.

\subsection{\label{cbpcme}Cartesian multipole expansion}
We are now interested in a Cartesian expansion for a function depending on the Euler angles. For this expansion, various options are possible. Ehrentraut and Muschik describe an expansion in higher-dimensional symmetric traceless tensors \cite{EhrentrautM1998}, which allows to construct order parameters for biaxial particles, since one can map between the sphere in four dimensions $S_3$ and the configuration space SO(3) of a biaxial particle \cite{EhrentrautM1998,JoslinG1984}. An orientational expansion of functions defined on $S_3$ can be performed in terms of four-dimensional spherical harmonics or, equivalently, higher-dimensional orientation vectors \cite{EhrentrautM1998}. A Cartesian expansion in four-dimensional symmetric traceless tensors is discussed in Ref.\ \cite{JoslinG1984}. 

We here describe an expansion derived by Turzi \cite{Turzi2011}. It allows to describe orientational order in terms of rotation matrices, which is geometrically more intuitive than using four-dimensional orientation vectors. Just as an angular expansion in terms of Wigner D-matrices can be obtained starting from the fact that they are representations of the rotation group SO(3), we can construct a Cartesian representation of this group to derive an expansion in outer products of a rotation matrix $R_{ij}$. The rotation matrix can be defined in terms of the Euler angles as \cite{GrayG1984}
\begin{widetext}
\begin{equation}
R(\vec{\varpi})=
\begin{pmatrix}
\cos(\phi)\cos(\theta)\cos(\chi) - \sin(\phi)\sin(\chi) & - \cos(\phi)\cos(\theta)\sin(\chi) - \sin(\phi)\cos(\chi) & \cos(\phi) \sin(\theta)\\
\sin(\phi)\cos(\theta)\cos(\chi) + \cos(\phi)\sin(\chi) & - \sin(\phi)\cos(\theta)\sin(\chi) + \cos(\phi)\cos(\chi) & \sin(\phi)\sin(\theta)\\
-\sin(\theta)\cos(\chi) & \sin(\theta)\sin(\chi) & \cos(\theta)\\
\end{pmatrix}.
\label{rotationmatrix}%
\end{equation}
\end{widetext}
This matrix describes a successive rotation as $R(\theta,\phi,\chi) = R_{z}(\phi)R_{y}(\theta)R_{z}(\chi)$ with the elementary rotation matrices
\begin{align}
&R_{y}(\varphi) = 
\begin{pmatrix}
\cos(\varphi)& 0&\sin(\varphi)\\
0& 1& 0\\
-\sin(\varphi)& 0& \cos(\varphi)\\
\end{pmatrix},\\
&R_{z}(\varphi) = 
\begin{pmatrix}
\cos(\varphi)& - \sin(\varphi) &0\\
\sin(\varphi)& \cos(\varphi)& 0\\
0 & 0& 1\\
\end{pmatrix}.
\end{align}
	
We denote by $(E^{(j)}_{-j},\dotsc,E^{(j)}_{j})$ an orthonormal basis of the vector space of symmetric traceless tensors of rank $j$. As the Wigner D-matrices for the angular case, an irreducible Cartesian representation $\tilde{D}(R)$ of the rotation group can be derived, which acts on a symmetric traceless tensor $T_{i_1 \dotsb i_l}$ as
\begin{equation}
(\tilde{D}(R)T)_{i_1\dotsb i_l} =
\!\! \sum_{j_1,\dotsc,j_l =1}^{3} \!\! R_{i_1 j_1}\!\dotsb R_{i_l j_l}T_{j_1 \dotsb j_l}.
\raisetag{1em}%
\end{equation}
Using the Peter-Weyl theorem, one can then expand a function $f \in L^2(\mathrm{SO(3)})$ in terms of the representation $\tilde{D}(R)$, whose matrix elements form a complete orthogonal system \cite{Turzi2011}. 
The resulting expansion is \cite{Turzi2011}
\begin{equation}
f(R) = \sum_{l=0}^{\infty} \sum_{i_{1},\dotsc,i_{l}=1}^{3} \sum_{j_{1},\dotsc,j_{l}=1}^{3} \!\!\!\!\! c^{(3\mathrm{D})}_{i_1 j_1 \dotsb i_l j_l}R_{i_1 j_1}\!\dotsb R_{i_l j_l}
\label{turzi}%
\end{equation}
with the expansion coefficients 
\begin{equation}
\begin{split}
c^{(3\mathrm{D})}_{i_1 j_1 \dotsb i_l j_l} &= \frac{2l+1}{8\pi^2}\INT{\mathrm{SO(3)}}{}{\Omega}\sum_{\alpha,\beta=-l}^{l}\sum_{a_{1},\dotsc,a_{l}=1}^{3}\sum_{b_{1},\dotsc,b_{l}=1}^{3}\\
&\quad\;\:\! (E^{(l)}_\alpha)_{a_1 \dotsb a_l}(E^{(l)}_\beta)_{b_1 \dotsb b_l}(E^{(l)}_\alpha)_{i_1 \dotsb i_l}(E^{(l)}_\beta)_{j_1 \dotsb j_l}\\
&\quad\;\:\! R_{a_1 b_1}\!\dotsb R_{a_l b_l}.
\end{split}\label{cIIID}\raisetag{3.6em}%
\end{equation}
In practical applications, the expansion coefficients can be calculated more efficiently in other ways (see Ref.\ \cite{Turzi2011} for details). A list of the biaxial expansion coefficients up to order $l=3$ can be found in \cref{tab:c}.
\begin{table*}[htbp]
\centering\renewcommand*{\arraystretch}{1.2}
\begin{tabularx}{\linewidth}{cY}
\hline
$\boldsymbol{l}$ & $\boldsymbol{\CIIIDIJa_{i_1 j_1 \dotsc i_l j_l}}$ \\\hline 
$0$ & $\frac{1}{8\pi^2}\TINT{\mathrm{SO(3)}}{}{\Omega} f(R)$ \\\hline 
$1$ & $\frac{3}{8\pi^2}\TINT{\mathrm{SO(3)}}{}{\Omega} f(R) R_{i_1j_1} $\\\hline 
$2$ & $\frac{5}{8\pi^2}\TINT{\mathrm{SO(3)}}{}{\Omega} f(R) (\frac{1}{2}(R_{i_1 j_1} R_{i_2 j_2} + R_{i_1 j_2} R_{i_2 j_1}) - \frac{1}{3}\delta_{i_1 i_2}\delta_{j_1 j_2})$  \\\hline $3$ & $\frac{7}{8\pi^2}\TINT{\mathrm{SO(3)}}{}{\Omega} f(R)(\frac{1}{6}(R_{i_1 j_1}R_{i_2 j_2}R_{i_3 j_3} + R_{i_1 j_2}R_{i_2 j_3}R_{i_3 j_1} + R_{i_1 j_3}R_{i_2 j_1}R_{i_3 j_2} + R_{i_1 j_2}R_{i_2 j_1}R_{i_3 j_3} + R_{i_1 j_1}R_{i_2 j_3}R_{i_3 j_2} +R_{i_1 j_3}R_{i_2 j_2}R_{i_3 j_1}) \newline 
\hspace*{2.4em}-\frac{1}{15}(R_{i_1 j_1}\delta_{i_2 i_3}\delta_{j_2 j_3} + R_{i_2 j_2}\delta_{i_1 i_3}\delta_{j_1 j_3} + R_{i_3 j_3}\delta_{i_1 i_2}\delta_{j_1 j_2} + R_{i_1 j_2}\delta_{i_2 i_3}\delta_{j_1 j_3} + R_{i_1 j_3}\delta_{i_2 i_3}\delta_{j_1 j_2} \newline 
\hspace*{-5.15em}+ R_{i_2 j_1}\delta_{i_1 i_3}\delta_{j_2 j_3} + R_{i_2 j_3}\delta_{i_1 i_3}\delta_{j_1 j_2} + R_{i_3 j_1}\delta_{i_1 i_2}\delta_{j_2 j_3} + R_{i_3 j_2}\delta_{i_1 i_2}\delta_{j_1 j_3} ))$  
\\\hline 
\raisebox{-0.6mm}[\ht\strutbox]{$\vdots$} & \raisebox{-0.6mm}[\ht\strutbox]{$\vdots$} \\\hline 
\end{tabularx}
\caption{\label{tab:c}Cartesian expansion coefficients $\CIIIDIJa_{i_1 j_1 \dotsc i_l j_l}$ for different orders $l$ \cite{Turzi2011}.}
\end{table*}

The expansion coefficients \eqref{cIIID} are symmetric and traceless in the $\{i_k\}$ and $\{j_k\}$ separately for $l>1$ (e.g., we have $c^{(3\mathrm{D})}_{1321} = c^{(3\mathrm{D})}_{2311}$ but $c^{(3\mathrm{D})}_{1321} \neq c^{(3\mathrm{D})}_{1231}$ and $\sum_{i=1}^{3} c^{(3\mathrm{D})}_{ikil}=0$ but $\sum_{i=1}^{3} c^{(3\mathrm{D})}_{iijl}\neq 0$), so the maximal number of independent expansion coefficients of a certain order $l$ is $(2l+1)^{2}$ and thus identical to that of the expansion coefficients \eqref{clmn} for the same $l$. In consequence, angular and Cartesian expansions are orderwise equivalent also in the biaxial case.

In \cref{relations}, explicit equations expressing the expansion coefficients of an angular multipole expansion in terms of the expansion coefficients of a Cartesian multipole expansion and vice versa are given up to third order for biaxial particles. These relations have been calculated using the same procedure as in the uniaxial case (see \cref{cartesianuniaxial}).

\subsection{Order parameters}
In the biaxial case, various definitions of orientational order parameters are possible and used in the literature. A very popular choice for $l=2$ is the Saup\'{e} ordering matrix, which can be used to describe outcomes of NMR measurements \cite{Rosso2007}. Here, we state the order parameters in the form that follows from the Turzi expansion for consistency with the previous section. 
We consider a system with a distribution function $f(\vec{r},R)$. A Cartesian expansion up to second order gives
\begin{equation}
\begin{split}%
f(\vec{r},R) = \rho(\vec{r}) + \! \SUM{i,j=1}{3}P_{ij}(\vec{r})R_{ij} + \!\!\!\!\! \SUM{i,j,k,l=1}{3}Q_{ijkl}(\vec{r})R_{ij}R_{kl} 
\raisetag{1.3em}%
\end{split}%
\end{equation}
with
\begin{align}
\rho(\vec{r}) &=\frac{1}{8\pi^2} \INT{\mathrm{SO(3)}}{}{\Omega}f(\vec{r},R),\\
P_{ij}(\vec{r}) &= \frac{3}{8\pi^2} \INT{\mathrm{SO(3)}}{}{\Omega}f(\vec{r},R)R_{ij},\\
\begin{split}
Q_{ijkl}(\vec{r}) &= \frac{5}{16\pi^2}\INT{\mathrm{SO(3)}}{}{\Omega}f(\vec{r},R)\Big(R_{ij}R_{kl} + R_{il}R_{kj} \\
&\qquad\qquad\qquad\qquad\quad\;\,\:\! - \frac{2}{3}\delta_{ik}\delta_{jl}\Big).
\end{split}\raisetag{3.0em}%
\end{align}
In particular, for the distribution function for a system of $N$ biaxial particles with positions $\vec{r}_n$ and orientations $R_n$ with $n=1,\dotsc,N$, given by
\begin{equation}
f(\vec{r},R) = \SUM{n=1}{N}\delta(\vec{r}-\vec{r}_n)\delta(R - R_n),
\end{equation}
we get the microscopic definitions
\begin{align}
\rho(\vec{r}) &= \frac{1}{8\pi^2}\sum_{n=1}^{N}\delta(\vec{r}-\vec{r}_n),\\
P_{ij}(\vec{r}) &= \frac{3}{8\pi^2}\sum_{n=1}^{N}R_{n,ij}\delta(\vec{r}-\vec{r}_n),\\
\begin{split}
Q_{ijkl}(\vec{r}) &= \frac{5}{16\pi^2}\sum_{n=1}^{N}\Big( R_{n,ij}R_{n,kl}  + R_{n,il}R_{n,kj} \\
&\qquad\qquad\qquad\! - \frac{2}{3}\delta_{ik}\delta_{jl}\Big)\delta(\vec{r}-\vec{r}_n)
\end{split}
\end{align}
with $R_{n,ij}=(R_n)_{ij}$ denoting an element of the matrix $R_n$.

It is counterintuitive that, in the biaxial case, the first- and second-rank tensors giving polarization and nematic order from the uniaxial case have to be replaced by a second- and fourth-rank tensor, respectively. One can understand this in analogy to quantum mechanics: For uniaxial particles, various orientational states are degenerate, such that one only needs to distinguish three rather than nine degrees of freedom at first order. This is similar as in the hydrogen atom, where various angular momentum eigenstates have the same energy and do not have to be distinguished in elementary treatments. In our case, the degeneracy is lifted by biaxiality.

There is also an interesting analogy between active and quantum systems. Passive spherical particles have no distinguishable orientations and thus no orientational order. However, active spherical particles have a preferred direction (the axis of self-propulsion), such that the orientational degeneracy is lifted. In consequence, systems of active spherical particles can exhibit orientational order. Staying in the comparison to the hydrogen atom, activity has a similar role as the magnetic field in the Zeeman effect, which lifts the degeneracy of the energy levels of the hydrogen atom \cite{Foot2005}.

\section{\label{conclusion}Conclusions}
In this article, we have discussed angular and Cartesian uniaxial and biaxial expansions in two and three spatial dimensions. We have given an overview over the relevant functions and definitions, and derived formulas for conversions between the expansion coefficients of both types of expansions up to third order. These formulas allow to relate the results of analytical calculations, computer simulations, and experiments based on different types of expansions, which makes it possible to combine their advantages by converting between them.

Possible continuations of this work include the consideration of angular and Cartesian expansions in higher-dimensional spaces, in particular concerning the relations between four-dimensional symmetric traceless tensors and the biaxial expansions discussed here. Moreover, the formalism could be extended towards vector and higher-order tensor spherical harmonics \cite{GrinterJ2018,Freeden2009,Fengler2005,Freeden1994} or quantum state multipoles \cite{Blum2012,delaHozBKLS2014}.

\section*{Acknowledgements}
We thank Jonas L\"ubken for helpful discussions. 
R.W.\ is funded by the Deutsche Forschungsgemeinschaft (DFG, German Research Foundation) -- WI 4170/3-1.

\appendix
\section{\label{relation}Relation of angular and Cartesian expansion coefficients for the uniaxial case}
The following equations allow to convert the expansion coefficients of an angular and Cartesian expansion, respectively, up to third order into each other. Higher-order relations can be derived using Eqs.\ \eqref{fphi}, \eqref{fk}, \eqref{fthetaphi}, \eqref{flm}, \eqref{fu}, and \eqref{eq:fdD}.
\begin{itemize}
\item \textit{Circular from Cartesian (2D):}\vspace{-2.6ex}
\end{itemize}
\begin{align}
f_{0}&=\ff{2}{0}{}, \\
f_{\pm 1}&=\frac{1}{2}(\ff{2}{1}{1}\mp\ii\ff{2}{1}{2}), \\
f_{\pm 2}&=\frac{1}{2}(\ff{2}{2}{11}\mp\ii\ff{2}{2}{12}), \\
f_{\pm 3}&=\frac{1}{2}(\ff{2}{3}{111}\mp\ii\ff{2}{3}{112})
\end{align}
\begin{itemize}
\item \textit{Cartesian from circular (2D):}\vspace{-2.6ex}
\end{itemize}
\begin{align}
\ff{2}{0}{}&=f_{0}, \\
\ff{2}{1}{1}&=f_{1}+f_{-1}, \\
\ff{2}{1}{2}&=\ii (f_{1}-f_{-1}), \\
\ff{2}{2}{11}&=f_{2}+f_{-2}, \\
\ff{2}{2}{12}&=\ff{2}{2}{21}=\ii (f_{2}-f_{-2}), \\
\ff{2}{2}{22}&=-\ff{2}{2}{11}, \\
\ff{2}{3}{111}&=f_{3}+f_{-3}, \\
\ff{2}{3}{112}&=\ff{2}{3}{121}=\ff{2}{3}{211}=\ii (f_{3}-f_{-3}), \\
\ff{2}{3}{122}&=\ff{2}{3}{212}=\ff{2}{3}{221}=-\ff{2}{3}{111}, \\
\ff{2}{3}{222}&=-\ff{2}{3}{211}
\end{align}
\begin{itemize}
\item \textit{Spherical from Cartesian (3D):}\vspace{-2.6ex}
\end{itemize}
\begin{align}
f_{00}&=2\sqrt{\pi}\ff{3}{0}{}, \\
f_{10}&=2\sqrt{\frac{\pi}{3}}\ff{3}{1}{3}, \\
f_{1\pm 1}&=\sqrt{\frac{2\pi}{3}}(\mp\ff{3}{1}{1}+\ii\ff{3}{1}{2}), \\
f_{20}&=2\sqrt{\frac{\pi}{5}}\ff{3}{2}{33}, \\
f_{2\pm 1}&=2\sqrt{\frac{2\pi}{15}}(\mp \ff{3}{2}{13}+\ii\ff{3}{2}{23}), \\
f_{2\pm 2}&=\sqrt{\frac{2\pi}{15}}(\ff{3}{2}{11}-\ff{3}{2}{22}\mp 2\ii\ff{3}{2}{12}), \\
f_{30}&=2\sqrt{\frac{\pi}{7}}\ff{3}{3}{333}, \\
f_{3\pm 1}&=\sqrt{\frac{3\pi}{7}}(\mp\ff{3}{3}{133}+\ii\ff{3}{3}{233}), \\
f_{3\pm 2}&=\sqrt{\frac{6\pi}{35}}(\ff{3}{3}{113}-\ff{3}{3}{223}\mp 2\ii\ff{3}{3}{123}), \\
f_{3\pm 3}&=\sqrt{\frac{\pi}{35}}(\mp\ff{3}{3}{111}\pm 3\ff{3}{3}{122}+3\ii\ff{3}{3}{112}-\ii\ff{3}{3}{222})
\end{align}
\begin{itemize}
\item \textit{Cartesian from spherical (3D):}\vspace{-2.6ex}
\end{itemize}
\begin{align}
\ff{3}{0}{}&=\frac{1}{2\sqrt{\pi}}f_{00}, \\
\ff{3}{1}{1}&=-\frac{1}{2}\sqrt{\frac{3}{2\pi}}(f_{11}-f_{1-1}), \\
\ff{3}{1}{2}&=-\ii\frac{1}{2}\sqrt{\frac{3}{2\pi}}(f_{11}+f_{1-1}), \\
\ff{3}{1}{3}&=\frac{1}{2}\sqrt{\frac{3}{\pi}}f_{10}, \\
\ff{3}{2}{11}&=-\frac{1}{8}\sqrt{\frac{5}{\pi}}(2f_{20}-\sqrt{6}(f_{22}+f_{2-2})), \\
\ff{3}{2}{12}&=\ff{3}{2}{21}=\ii\frac{1}{4}\sqrt{\frac{15}{2\pi}}(f_{22}-f_{2-2}), \\
\ff{3}{2}{13}&=\ff{3}{2}{31}=-\frac{1}{4}\sqrt{\frac{15}{2\pi}}(f_{21}-f_{2-1}), \\
\ff{3}{2}{22}&=-\frac{1}{8}\sqrt{\frac{5}{\pi}}(2f_{20}+\sqrt{6}(f_{22}+f_{2-2})), \\
\ff{3}{2}{23}&=\ff{3}{2}{32}=-\ii\frac{1}{4}\sqrt{\frac{15}{2\pi}}(f_{21}+f_{2-1}), \\
\ff{3}{2}{33}&=-\ff{3}{2}{11}-\ff{3}{2}{22}, \\
\ff{3}{3}{111}&=\frac{1}{8}\sqrt{\frac{7}{\pi}}(\sqrt{3}(f_{31}-f_{3-1})-\sqrt{5}(f_{33}-f_{3-3})), \raisetag{2ex}\\
\begin{split}
\ff{3}{3}{112}&=\ff{3}{3}{121}=\ff{3}{3}{211} \\
&=\ii\frac{1}{24}\sqrt{\frac{21}{\pi}}(f_{31}+f_{3-1}-\sqrt{15}(f_{33}+f_{3-3})), 
\end{split}\raisetag{7ex}\\
\begin{split}
\ff{3}{3}{113}&=\ff{3}{3}{131}=\ff{3}{3}{311} \\
&=-\frac{1}{24}\sqrt{\frac{42}{\pi}}(\sqrt{6}f_{30}-\sqrt{5}(f_{32}+f_{3-2})), 
\end{split}\\
\begin{split}
\ff{3}{3}{122}&=\ff{3}{3}{212}=\ff{3}{3}{221}\\
&=\frac{1}{24}\sqrt{\frac{21}{\pi}}(f_{31}-f_{3-1}+\sqrt{15}(f_{33}-f_{3-3})), 
\end{split}\\
\begin{split}
\ff{3}{3}{123}&=\ff{3}{3}{132}=\ff{3}{3}{213}=\ff{3}{3}{231}=\ff{3}{3}{312}\\
&=\ff{3}{3}{321}=\ii\frac{1}{4}\sqrt{\frac{35}{6\pi}}(f_{32}-f_{3-2}), 
\end{split}\\
\ff{3}{3}{133}&=\ff{3}{3}{313}=\ff{3}{3}{331}=-\ff{3}{3}{111}-\ff{3}{3}{122}, \\
\ff{3}{3}{222}&=\ii\frac{1}{8}\sqrt{\frac{7}{\pi}}(\sqrt{3}(f_{31}+f_{3-1})+\sqrt{5}(f_{33}+f_{3-3})), \raisetag{2ex}\\
\begin{split}
\ff{3}{3}{223}&=\ff{3}{3}{232}=\ff{3}{3}{322} \\
&=-\frac{1}{24}\sqrt{\frac{42}{\pi}}(\sqrt{6}f_{30}+\sqrt{5}(f_{32}+f_{3-2})), 
\end{split}\\
\ff{3}{3}{233}&=\ff{3}{3}{323}=\ff{3}{3}{332}=-\ff{3}{3}{211}-\ff{3}{3}{222}, \\
\ff{3}{3}{333}&=-\ff{3}{3}{311}-\ff{3}{3}{322} 
\end{align}

\section{\label{relations}Relation of angular and Cartesian expansion coefficients for the biaxial case}
With the following equations, the expansion coefficients of an angular and Cartesian expansion, respectively, up to third order can be converted into each other. Relations for higher orders can be derived using Eqs.\ \eqref{dexp}, \eqref{clmn}, \eqref{turzi}, and \eqref{cIIID}.

\begin{itemize}
\item \textit{Angular from Cartesian:}\vspace{-2.6ex}
\end{itemize}

\begin{itemize}
\item \textit{Cartesian from angular:}\vspace{-2.6ex}
\end{itemize}
%
The expansion coefficients that appear not explicitly in the previous table follow from their symmetry properties
\begin{align}
\begin{split}\CIIIDIJc{i_{1}}{j_{1}}{i_{2}}{j_{2}}&=\CIIIDIJc{i_{2}}{j_{1}}{i_{1}}{j_{2}}=\CIIIDIJc{i_{1}}{j_{2}}{i_{2}}{j_{1}},\end{split}\\
\begin{split}\CIIIDIJd{i_{1}}{j_{1}}{i_{2}}{j_{2}}{i_{3}}{j_{3}}&=\CIIIDIJd{i_{2}}{j_{1}}{i_{1}}{j_{2}}{i_{3}}{j_{3}}=\CIIIDIJd{i_{3}}{j_{1}}{i_{2}}{j_{2}}{i_{1}}{j_{3}}\\
&=\CIIIDIJd{i_{1}}{j_{1}}{i_{3}}{j_{2}}{i_{2}}{j_{3}}=\CIIIDIJd{i_{1}}{j_{2}}{i_{2}}{j_{1}}{i_{3}}{j_{3}}\\
&=\CIIIDIJd{i_{1}}{j_{3}}{i_{2}}{j_{2}}{i_{3}}{j_{1}}=\CIIIDIJd{i_{1}}{j_{1}}{i_{2}}{j_{3}}{i_{3}}{j_{2}}\end{split}
\end{align}
and tracelessness 
\begin{align}
\begin{split}\CIIIDIJc{3}{j_{1}}{3}{j_{2}}  &=  -\CIIIDIJc{1}{j_{1}}{1}{j_{2}} - \CIIIDIJc{2}{j_{1}}{2}{j_{2}},\end{split}\\
\begin{split}\CIIIDIJc{i_{1}}{3}{i_{2}}{3}  &=  -\CIIIDIJc{i_{1}}{1}{i_{2}}{1} - \CIIIDIJc{i_{1}}{2}{i_{2}}{2},\end{split}\\
\begin{split}\CIIIDIJd{3}{j_{1}}{3}{j_{2}}{i_{3}}{j_{3}}  &=  -\CIIIDIJd{1}{j_{1}}{1}{j_{2}}{i_{3}}{j_{3}} - \CIIIDIJd{2}{j_{1}}{2}{j_{2}}{i_{3}}{j_{3}},\end{split}\\
\begin{split}\CIIIDIJd{3}{j_{1}}{i_{2}}{j_{2}}{3}{j_{3}}  &=  -\CIIIDIJd{1}{j_{1}}{i_{2}}{j_{2}}{1}{j_{3}} - \CIIIDIJd{2}{j_{1}}{i_{2}}{j_{2}}{2}{j_{3}},\end{split}\\
\begin{split}\CIIIDIJd{i_{1}}{j_{1}}{3}{j_{2}}{3}{j_{3}}  &=  -\CIIIDIJd{i_{1}}{j_{1}}{1}{j_{2}}{1}{j_{3}} - \CIIIDIJd{i_{1}}{j_{1}}{2}{j_{2}}{2}{j_{3}},\end{split}\\
\begin{split}\CIIIDIJd{i_{1}}{3}{i_{2}}{3}{i_{3}}{j_{3}}  &=  -\CIIIDIJd{i_{1}}{1}{i_{2}}{1}{i_{3}}{j_{3}} - \CIIIDIJd{i_{1}}{2}{i_{2}}{2}{i_{3}}{j_{3}},\end{split}\\
\begin{split}\CIIIDIJd{i_{1}}{3}{i_{2}}{j_{2}}{i_{3}}{3}  &=  -\CIIIDIJd{i_{1}}{1}{i_{2}}{j_{2}}{i_{3}}{1} - \CIIIDIJd{i_{1}}{2}{i_{2}}{j_{2}}{i_{3}}{2},\end{split}\\
\begin{split}\CIIIDIJd{i_{1}}{j_{1}}{i_{2}}{3}{i_{3}}{3}  &=  -\CIIIDIJd{i_{1}}{j_{1}}{i_{2}}{1}{i_{3}}{1} - \CIIIDIJd{i_{1}}{j_{1}}{i_{2}}{2}{i_{3}}{2}\end{split}.
\end{align}

\section{\label{list}Elements of Wigner D-matrices}
Finally, we list all elements of the Wigner D-matrices \eqref{eq:WignerDmatrices} with $l \leq 3$ \cite{GrayG1984,VarshalovichMK1988}: 
{\allowdisplaybreaks\begin{align}
\begin{split}\WDvarpi{0}{0}{0} &= 1,\end{split}\\
\begin{split}\WDvarpi{1}{-1}{-1} &= e^{\ii \phi} \frac{1}{2}(1+\cos(\theta)) e^{\ii \chi},\end{split}\\
\begin{split}\WDvarpi{1}{-1}{0} &= e^{\ii \phi} \frac{\sin(\theta)}{\sqrt{2}},\end{split}\\
\begin{split}\WDvarpi{1}{-1}{1} &= e^{\ii \phi} \frac{1}{2}(1-\cos(\theta)) e^{-\ii \chi},\end{split}\\
\begin{split}\WDvarpi{1}{0}{-1} &= -\frac{\sin(\theta)}{\sqrt{2}} e^{\ii \chi},\end{split}\\
\begin{split}\WDvarpi{1}{0}{0} &= \cos(\theta),\end{split}\\
\begin{split}\WDvarpi{1}{0}{1} &= \frac{\sin(\theta)}{\sqrt{2}} e^{-\ii \chi},\end{split}\\
\begin{split}\WDvarpi{1}{1}{-1} &= e^{-\ii \phi} \frac{1}{2}(1-\cos(\theta)) e^{\ii \chi},\end{split}\\
\begin{split}\WDvarpi{1}{1}{0} &= -e^{-\ii \phi} \frac{\sin(\theta)}{\sqrt{2}},\end{split}\\
\begin{split}\WDvarpi{1}{1}{1} &= e^{-\ii \phi} \frac{1}{2}(1+\cos(\theta)) e^{-\ii \chi},\end{split}\\
\begin{split}\WDvarpi{2}{-2}{-2} &= e^{2 \ii \phi} \frac{1}{4}(1+\cos(\theta))^2 e^{2 \ii \chi},\end{split}\\
\begin{split}\WDvarpi{2}{-2}{-1} &= e^{2 \ii \phi} \frac{1}{2} \sin(\theta)(1+\cos(\theta)) e^{\ii \chi},\end{split}\\
\begin{split}\WDvarpi{2}{-2}{0} &= e^{2 \ii \phi} \frac{1}{2} \sqrt{\frac{3}{2}} \sin(\theta)^{2},\end{split}\\
\begin{split}\WDvarpi{2}{-2}{1} &= -e^{2 \ii \phi} \frac{1}{2} \sin(\theta)(\cos(\theta)-1) e^{-\ii \chi},\end{split}\\
\begin{split}\WDvarpi{2}{-2}{2} &= e^{2 \ii \phi} \frac{1}{4}(\cos(\theta)-1)^2 e^{-2 \ii \chi},\end{split}\\
\begin{split}\WDvarpi{2}{-1}{-2} &= -e^{\ii \phi} \frac{1}{2} \sin(\theta)(1+\cos(\theta)) e^{2 \ii \chi},\end{split}\\
\begin{split}\WDvarpi{2}{-1}{-1} &= e^{\ii \phi} \frac{1}{2}(1+\cos(\theta))(2\cos(\theta)-1) e^{\ii \chi},\end{split}\\
\begin{split}\WDvarpi{2}{-1}{0} &= e^{\ii \phi} \sqrt{\frac{3}{2}} \sin(\theta) \cos(\theta),\end{split}\\
\begin{split}\WDvarpi{2}{-1}{1} &= e^{\ii \phi} \frac{1}{2}(1-\cos(\theta))(1+2\cos(\theta)) e^{-\ii \chi},\end{split}\\
\begin{split}\WDvarpi{2}{-1}{2} &= -e^{\ii \phi} \frac{1}{2} \sin(\theta)(\cos(\theta)-1) e^{-2 \ii \chi},\end{split}\\
\begin{split}\WDvarpi{2}{0}{-2} &= \frac{1}{2} \sqrt{\frac{3}{2}} \sin(\theta)^{2} e^{2 \ii \chi},\end{split}\\
\begin{split}\WDvarpi{2}{0}{-1} &= -\sqrt{\frac{3}{2}} \sin(\theta) \cos(\theta) e^{\ii \chi},\end{split}\\
\begin{split}\WDvarpi{2}{0}{0} &= \frac{1}{2}(3 \cos(\theta)^{2}-1),\end{split}\\
\begin{split}\WDvarpi{2}{0}{1} &= \sqrt{\frac{3}{2}} \sin(\theta) \cos(\theta) e^{-\ii \chi},\end{split}\\
\begin{split}\WDvarpi{2}{0}{2} &= \frac{1}{2} \sqrt{\frac{3}{2}} \sin(\theta)^{2} e^{-2 \ii \chi},\end{split}\\
\begin{split}\WDvarpi{2}{1}{-2} &= e^{-\ii \phi} \frac{1}{2} \sin(\theta)(\cos(\theta)-1) e^{2 \ii \chi},\end{split}\\
\begin{split}\WDvarpi{2}{1}{-1} &= e^{-\ii \phi} \frac{1}{2}(1-\cos(\theta))(1+2\cos(\theta)) e^{\ii \chi},\end{split}\\
\begin{split}\WDvarpi{2}{1}{0} &= -e^{-\ii \phi} \sqrt{\frac{3}{2}} \sin(\theta) \cos(\theta),\end{split}\\
\begin{split}\WDvarpi{2}{1}{1} &= e^{-\ii \phi} \frac{1}{2}(1+\cos(\theta))(2\cos(\theta)-1) e^{-\ii \chi},\end{split}\\
\begin{split}\WDvarpi{2}{1}{2} &= e^{-\ii \phi} \frac{1}{2} \sin(\theta)(1+\cos(\theta)) e^{-2 \ii \chi},\end{split}\\
\begin{split}\WDvarpi{2}{2}{-2} &= e^{-2 \ii \phi} \frac{1}{4}(\cos(\theta)-1)^2 e^{2 \ii \chi},\end{split}\\
\begin{split}\WDvarpi{2}{2}{-1} &= e^{-2 \ii \phi} \frac{1}{2} \sin(\theta)(\cos(\theta)-1) e^{\ii \chi},\end{split}\\
\begin{split}\WDvarpi{2}{2}{0} &= e^{-2 \ii \phi} \frac{1}{2} \sqrt{\frac{3}{2}} \sin(\theta)^{2},\end{split}\\
\begin{split}\WDvarpi{2}{2}{1} &= -e^{-2 \ii \phi} \frac{1}{2} \sin(\theta)(1+\cos(\theta)) e^{-\ii \chi},\end{split}\\
\begin{split}\WDvarpi{2}{2}{2} &= e^{-2 \ii \phi} \frac{1}{4}(1+\cos(\theta))^2 e^{-2 \ii \chi},\end{split}\\
\begin{split}\WDvarpi{3}{-3}{-3} &= e^{3 \ii \phi} \frac{1}{8}(1+\cos(\theta))^3 e^{3 \ii \chi},\end{split}\\
\begin{split}\WDvarpi{3}{-3}{-2} &= e^{3 \ii \phi} \frac{1}{4} \sqrt{\frac{3}{2}} \sin(\theta)(1+\cos(\theta))^2 e^{2 \ii \chi},\end{split}\\
\begin{split}\WDvarpi{3}{-3}{-1} &= e^{3 \ii \phi} \frac{1}{8} \sqrt{15} \sin(\theta)^{2}(1+\cos(\theta)) e^{\ii \chi},\end{split}\\
\begin{split}\WDvarpi{3}{-3}{0} &= e^{3 \ii \phi} \frac{1}{4} \sqrt{5} \sin(\theta)^{3},\end{split}\\
\begin{split}\WDvarpi{3}{-3}{1} &= -e^{3 \ii \phi} \frac{1}{8} \sqrt{15} \sin(\theta)^{2}(\cos(\theta)-1) e^{-\ii \chi},\end{split}\\
\begin{split}\WDvarpi{3}{-3}{2} &= e^{3 \ii \phi} \frac{1}{4} \sqrt{\frac{3}{2}} \sin(\theta)(\cos(\theta)-1)^2 e^{-2 \ii \chi},\end{split}\\
\begin{split}\WDvarpi{3}{-3}{3} &= -e^{3 \ii \phi} \frac{1}{8}(\cos(\theta)-1)^3 e^{-3 \ii \chi},\end{split}\\
\begin{split}\WDvarpi{3}{-2}{-3} &= -e^{2 \ii \phi} \frac{1}{4} \sqrt{\frac{3}{2}} \sin(\theta)(1+\cos(\theta))^2 e^{3 \ii \chi},\end{split}\\
\begin{split}\WDvarpi{3}{-2}{-2} &= e^{2 \ii \phi} \frac{1}{4}(1+\cos(\theta))^2(3 \cos(\theta)-2) e^{2 \ii \chi},\end{split}\\
\begin{split}\WDvarpi{3}{-2}{-1} &= e^{2 \ii \phi} \frac{1}{4} \sqrt{\frac{5}{2}} \sin(\theta)(1+\cos(\theta))\\
&\EinrueckabstandII (3 \cos(\theta)-1) e^{\ii \chi},\end{split}\\
\begin{split}\WDvarpi{3}{-2}{0} &= e^{2 \ii \phi} \frac{1}{2} \sqrt{\frac{15}{2}} \sin(\theta)^{2} \cos(\theta),\end{split}\\
\begin{split}\WDvarpi{3}{-2}{1} &= e^{2 \ii \phi} \frac{1}{4} \sqrt{\frac{5}{2}} \sin(\theta)(1-\cos(\theta))\\
&\EinrueckabstandII (1+3\cos(\theta)) e^{-\ii \chi},\end{split}\\
\begin{split}\WDvarpi{3}{-2}{2} &= e^{2 \ii \phi} \frac{1}{4}(\cos(\theta)-1)^2(2+3\cos(\theta)) e^{-2 \ii \chi},\end{split}\\
\begin{split}\WDvarpi{3}{-2}{3} &= e^{2 \ii \phi} \frac{1}{4} \sqrt{\frac{3}{2}} \sin(\theta)(\cos(\theta)-1)^2 e^{-3 \ii \chi},\end{split}\\
\begin{split}\WDvarpi{3}{-1}{-3} &= e^{\ii \phi} \frac{1}{8} \sqrt{15} \sin(\theta)^{2}(1+\cos(\theta)) e^{3 \ii \chi},\end{split}\\
\begin{split}\WDvarpi{3}{-1}{-2} &= -e^{\ii \phi} \frac{1}{4} \sqrt{\frac{5}{2}} \sin(\theta)(1+\cos(\theta))\\
&\EinrueckabstandII (3 \cos(\theta)-1) e^{2 \ii \chi},\end{split}\\
\begin{split}\WDvarpi{3}{-1}{-1} &= e^{\ii \phi} \frac{1}{8}(1+\cos(\theta))(15\cos(\theta)^{2}\\
&\EinrueckabstandI -10\cos(\theta)-1) e^{\ii \chi},\end{split}\\
\begin{split}\WDvarpi{3}{-1}{0} &= e^{\ii \phi} \frac{1}{4} \sqrt{3} \sin(\theta)(5 \cos(\theta)^{2}-1),\end{split}\\
\begin{split}\WDvarpi{3}{-1}{1} &= e^{\ii \phi} \frac{1}{8}(1-\cos(\theta))(15\cos(\theta)^{2}\\
&\EinrueckabstandI +10\cos(\theta)-1) e^{-\ii \chi},\end{split}\\
\begin{split}\WDvarpi{3}{-1}{2} &= e^{\ii \phi} \frac{1}{4} \sqrt{\frac{5}{2}} \sin(\theta)(1-\cos(\theta))\\
&\EinrueckabstandII (1+3\cos(\theta)) e^{-2 \ii \chi},\end{split}\\
\begin{split}\WDvarpi{3}{-1}{3} &= -e^{\ii \phi} \frac{1}{8} \sqrt{15} \sin(\theta)^{2}(\cos(\theta)-1) e^{-3 \ii \chi},\end{split}\\
\begin{split}\WDvarpi{3}{0}{-3} &= -\frac{1}{4} \sqrt{5} \sin(\theta)^{3} e^{3 \ii \chi},\end{split}\\
\begin{split}\WDvarpi{3}{0}{-2} &= \frac{1}{2} \sqrt{\frac{15}{2}} \sin(\theta)^{2} \cos(\theta) e^{2 \ii \chi},\end{split}\\
\begin{split}\WDvarpi{3}{0}{-1} &= \frac{1}{4} \sqrt{3} \sin(\theta)(1-5 \cos(\theta)^{2}) e^{\ii \chi},\end{split}\\
\begin{split}\WDvarpi{3}{0}{0} &= \frac{1}{2} \cos(\theta)(5 \cos(\theta)^{2}-3),\end{split}\\
\begin{split}\WDvarpi{3}{0}{1} &= \frac{1}{4} \sqrt{3} \sin(\theta)(5 \cos(\theta)^{2}-1) e^{-\ii \chi},\end{split}\\
\begin{split}\WDvarpi{3}{0}{2} &= \frac{1}{2} \sqrt{\frac{15}{2}} \sin(\theta)^{2} \cos(\theta) e^{-2 \ii \chi},\end{split}\\
\begin{split}\WDvarpi{3}{0}{3} &= \frac{1}{4} \sqrt{5} \sin(\theta)^{3} e^{-3 \ii \chi},\end{split}\\
\begin{split}\WDvarpi{3}{1}{-3} &= -e^{-\ii \phi} \frac{1}{8} \sqrt{15} \sin(\theta)^{2}(\cos(\theta)-1) e^{3 \ii \chi},\end{split}\\
\begin{split}\WDvarpi{3}{1}{-2} &= e^{-\ii \phi} \frac{1}{4} \sqrt{\frac{5}{2}} \sin(\theta)(\cos(\theta)-1)\\
&\EinrueckabstandII (1+3\cos(\theta)) e^{2 \ii \chi},\end{split}\\
\begin{split}\WDvarpi{3}{1}{-1} &= e^{-\ii \phi} \frac{1}{8}(1-\cos(\theta))(15\cos(\theta)^{2}\\
&\EinrueckabstandI +10\cos(\theta)-1) e^{\ii \chi},\end{split}\\
\begin{split}\WDvarpi{3}{1}{0} &= e^{-\ii \phi} \frac{1}{4} \sqrt{3} \sin(\theta)(1-5 \cos(\theta)^{2}),\end{split}\\
\begin{split}\WDvarpi{3}{1}{1} &= e^{-\ii \phi} \frac{1}{8}(1+\cos(\theta))(15\cos(\theta)^{2}\\
&\EinrueckabstandI-10\cos(\theta)-1) e^{-\ii \chi},\end{split}\\
\begin{split}\WDvarpi{3}{1}{2} &= e^{-\ii \phi} \frac{1}{4} \sqrt{\frac{5}{2}} \sin(\theta)(1+\cos(\theta))\\
&\EinrueckabstandII (3 \cos(\theta)-1) e^{-2 \ii \chi},\end{split}\\
\begin{split}\WDvarpi{3}{1}{3} &= e^{-\ii \phi} \frac{1}{8} \sqrt{15} \sin(\theta)^{2}(1+\cos(\theta)) e^{-3 \ii \chi},\end{split}\\
\begin{split}\WDvarpi{3}{2}{-3} &= -e^{-2 \ii \phi} \frac{1}{4} \sqrt{\frac{3}{2}} \sin(\theta)(\cos(\theta)-1)^2 e^{3 \ii \chi},\end{split}\\
\begin{split}\WDvarpi{3}{2}{-2} &= e^{-2 \ii \phi} \frac{1}{4}(\cos(\theta)-1)^2(2+3\cos(\theta)) e^{2 \ii \chi},\end{split}\\
\begin{split}\WDvarpi{3}{2}{-1} &= e^{-2 \ii \phi} \frac{1}{4} \sqrt{\frac{5}{2}} \sin(\theta)(\cos(\theta)-1)\\
&\EinrueckabstandII (1+3\cos(\theta)) e^{\ii \chi},\end{split}\\
\begin{split}\WDvarpi{3}{2}{0} &= e^{-2 \ii \phi} \frac{1}{2} \sqrt{\frac{15}{2}} \sin(\theta)^{2} \cos(\theta),\end{split}\\
\begin{split}\WDvarpi{3}{2}{1} &= -e^{-2 \ii \phi} \frac{1}{4} \sqrt{\frac{5}{2}} \sin(\theta)(1+\cos(\theta))\\
&\EinrueckabstandII (3 \cos(\theta)-1) e^{-\ii \chi},\end{split}\\
\begin{split}\WDvarpi{3}{2}{2} &= e^{-2 \ii \phi} \frac{1}{4}(1+\cos(\theta))^2(3 \cos(\theta)-2)\\
&\EinrueckabstandII e^{-2 \ii \chi},\end{split}\\
\begin{split}\WDvarpi{3}{2}{3} &= e^{-2 \ii \phi} \frac{1}{4} \sqrt{\frac{3}{2}} \sin(\theta)(1+\cos(\theta))^2 e^{-3 \ii \chi},\end{split}\\
\begin{split}\WDvarpi{3}{3}{-3} &= -e^{-3 \ii \phi} \frac{1}{8}(\cos(\theta)-1)^3 e^{3 \ii \chi},\end{split}\\
\begin{split}\WDvarpi{3}{3}{-2} &= -e^{-3 \ii \phi} \frac{1}{4} \sqrt{\frac{3}{2}} \sin(\theta)(\cos(\theta)-1)^2 e^{2 \ii \chi},\end{split}\\
\begin{split}\WDvarpi{3}{3}{-1} &= -e^{-3 \ii \phi} \frac{1}{8} \sqrt{15} \sin(\theta)^{2}(\cos(\theta)-1) e^{\ii \chi},\end{split}\\
\begin{split}\WDvarpi{3}{3}{0} &= -e^{-3 \ii \phi} \frac{1}{4} \sqrt{5} \sin(\theta)^{3},\end{split}\\
\begin{split}\WDvarpi{3}{3}{1} &= e^{-3 \ii \phi} \frac{1}{8} \sqrt{15} \sin(\theta)^{2}(1+\cos(\theta)) e^{-\ii \chi},\end{split}\\
\begin{split}\WDvarpi{3}{3}{2} &= -e^{-3 \ii \phi} \frac{1}{4} \sqrt{\frac{3}{2}} \sin(\theta)(1+\cos(\theta))^2 e^{-2 \ii \chi},\end{split}\\
\begin{split}\WDvarpi{3}{3}{3} &= e^{-3 \ii \phi} \frac{1}{8}(1+\cos(\theta))^3 e^{-3 \ii \chi}\end{split}
\end{align}}

\nocite{apsrev41Control}
\bibliographystyle{apsrev4-1}
\bibliography{control,refs}

\end{document}